\documentclass[aps,pre,twocolumn,amsmath,amssymb,showpacs,amsfonts]{revtex4}
\usepackage{epsfig}
\usepackage{graphicx}
\usepackage{dcolumn}
\usepackage{bm}

\begin{document}

\title{Exact Scaling Relations In Relativistic Hydrodynamic Turbulence}

\author{Itzhak Fouxon}
\author{Yaron Oz}

\affiliation{Raymond and Beverly Sackler School of Physics and Astronomy,
Tel-Aviv University, Tel-Aviv 69978, Israel}

\date{\today}

\begin{abstract}

We consider the steady state statistics of
turbulence in general classes of dissipative hydrodynamic equations, where the fluctuations are sustained by a random source concentrated at large scales.
It is well known that in some particular cases,
such as non-relativistic incompressible turbulence, a Kolmogorov-type exact scaling relation for a correlation function holds.
We show that all such scaling relations follow from a general relation on the current-density correlation
function.
The derivation
does not require an energy cascade picture and suggests that this traditional interpretation
of the Kolmogorov relation for incompressible turbulence may be misleading.
Using this we derive exact scaling results for compressible turbulence in relativistic hydrodynamics, which reduce
in the slow motion limit to the Kolmogorov relation.
We discuss the experimental implications of the results.

\end{abstract}


\maketitle

Developed incompressible turbulence of fluids is one of the oldest problems of theoretical physics. Very little
success has been achieved in the area due to the insurmountable difficulties presented by the strong non-linearity. Essentially, the only
non-trivial exact result on the statistics was derived in $1941$ by Kolmogorov \cite{Kolmogorov,Frisch,Gawedzki}. He derived an exact relation
stating that the third order correlation function of the flow velocity scales linearly with
the distance in a wide range of scales called the inertial interval. This relation is often interpreted as a statement
on the energy cascade: the energy pumped into the system by a source at large scales cascades downscales due to flow instabilities. Eventually
the energy is dissipated at a small scale by viscosity. In the wide intermediate range of scales, called inertial,
viscous dissipation is negligible, the energy is conserved and is merely being passed on to smaller-scale fluctuations.
Kolmogorov's relation expresses the constancy of the mean energy flux in the inertial range.
The inertial range statistics is of much interest due to the conjecture that some properties of the statistics are universal
in this range of scales. In particular, one expects a non-trivial scaling of the velocity correlation functions in the inertial range and the breakdown
of self-similarity \cite{Frisch,FalkovichSreenivasan}.
Currently, the scaling problem seems inaccessible and the
Kolmogorov relation is the only solid result.

While it is natural to approach any problem of a turbulent type by first deriving a Kolmogorov-type relation for it, this
was done actually only for quite a few problems. These are the passive scalar turbulence \cite{Yaglom}, the magnetohydrodynamic
turbulence \cite{Chandrasekhar,PouquetPolitano} and the Hall magnetohydrodynamics \cite{Galtier}. An important example of a
situation, where to the best of authors' knowledge no Kolmogorov-type relation has been derived is provided by the compressible
turbulence. In this letter we describe a general relation for steady states of a turbulent type. This relation reduces to known
Kolmogorov-type relations in the situations listed above, while in other situations it gives new results.
Our analysis indicates that the interpretation of the Kolmogorov relation for the incompressible turbulence in terms of the
energy cascade may be misleading (see also \cite{FFO}).
%In fact, the relation can be found from the condition of stationarity of the pair-correlation
%function of velocity in the steady state, see e. g. \cite{Gawedzki} and below, and no energy cascade picture is needed for the
%derivation.

In this Letter we derive exact scaling results for compressible turbulence in relativistic hydrodynamics, which reduce
in the slow motion limit to the Kolmogorov relation.
We discuss the implications of the results in view of the possible applications to the description of the hydrodynamic behavior of the quark-gluon plasma and to
condensed matter
physics, as well as due to
the existence of a dual gravitational description in terms of black hole geometry in asymptotically anti-de-Sitter space.

We consider a general class of classical field dynamics,
\begin{eqnarray}&&
\partial_t q^a+\nabla\cdot \bm j^a=f^a,
\end{eqnarray}
where $q^a, a=1,...,N$ are charges, $\bm j^a$ are currents and $f^a$ are the external random source fields.
These equations describe local conservation laws and  provide a canonical form for the effective hydrodynamic description.
The latter studies the evolution of slow modes of the system provided naturally by the local conservation laws
(since the zero wave-number component of charge fields is conserved, the low wave-number components evolve slowly by continuity  \cite{Forster}).
The equations are closed via a constitutive relation that expresses currents in terms of the charges. In accord with the fact
that hydrodynamics constitutes a low energy (wave-number) approximation, the relation has the
form of a series in gradients,
\begin{eqnarray}&&
j_i^a=F^a_i(\{q\})+\sum_{jb}G^a_{i, jb}(\{q\})\nabla_j q^b+\ldots, \label{exp}
\end{eqnarray}
where dots stand for higher order terms involving either higher order derivatives or first order derivatives in power larger than one.
Normally the zeroth order, reactive, term leads to a conservative dynamics while the first order term describes dissipation \cite{Forster}.
For our purposes here the consideration of higher order terms in Eq.~(\ref{exp}) is unnecessary and we will limit ourselves to the
general class of dynamics of the type,
\begin{eqnarray}&&
\partial_t q^a+\frac{\partial F_i^a}{\partial r_i}=f^a-\frac{\partial }{\partial r_i}\left(\sum_{jb}G^a_{i, jb}(\{q\})\nabla_j q^b
\right). \label{dyn}
\end{eqnarray}
The above truncation is usually sufficient for practical purposes. We will assume the standard mathematical formulation of the problem
of turbulence where the forcing term $f^a$ is random and its statistics is stationary, spatially homogeneous and isotropic. This implies
 that the same
properties of statistics hold for the steady state statistics of $q^a$. The correlation length of the force will be denoted below
by $L$.

An important example of the above class is provided by the
incompressible Navier-Stokes equations,
\begin{eqnarray}&&
\nabla\cdot\bm v=0,\ \ \partial_t \bm v+(\bm v\cdot\nabla)\bm v=-\nabla p+\nu\nabla^2 \bm v+\bm f \ , \label{b4}
\end{eqnarray}
where $\bm v$ is the flow velocity, $p$ is the pressure, $\nu$ is the kinematic viscosity and $f$ is an external force
field.
It is for these equations that the Kolmogorov relation was derived originally \cite{Kolmogorov}.
Consider the following version of the derivation,
where one
considers the steady state condition $\partial_t \langle v_i(0, t)v_i(\bm r, t)\rangle=0$, and the
angular brackets stand for averaging over the force statistics.
Using homogeneity and isotropy it follows from Eq.~(\ref{b4})
\begin{eqnarray}&&
0=\partial_t \langle v_i(0, t)v_i(\bm r, t)\rangle=-2\partial_j\langle v_i(0, t)v_i(\bm r, t)v_j(\bm r, t)\rangle
\nonumber\\&&+
2\langle v_i(0, t)f_i(\bm r, t)\rangle+2\nu \nabla^2 \langle v_i(0, t)v_i(\bm r, t)\rangle.
\end{eqnarray}
We now consider the limit of large correlation length $L$ of the forcing. This allows us to consider the limit of $r$ large but obeying
$r\ll L$.  Due to $r\ll L$ we have $f_i(\bm r, t)\approx f_i(0, t)$ and $\langle v_i(0, t)f_i(\bm r, t)\rangle\approx \langle v_i(0, t)f_i(0, t)\rangle\equiv\epsilon$ where $\epsilon$ is some constant equal in this case to the energy pumping rate. At sufficiently large $r$ the
last term in the above equation is negligible, since it contains a higher number of spatial derivatives, and we obtain
\begin{eqnarray}&&
\partial_j\langle v_i(0, t)v_i(\bm r, t)v_j(\bm r, t)\rangle=\epsilon,\ \\&& \langle v_i(0, t)v_i(\bm r, t)v_j(\bm r, t)\rangle=\epsilon r_j/d,
\label{Kolmogorov}
\end{eqnarray}
where $d$ is the spatial dimension and we used isotropy. The above relation on the triple correlation function is the Kolmogorov relation
and using statistical symmetries it can be brought to the standard form
\begin{equation}
\langle [(\bm v(\bm r)-\bm v(0))\cdot \bm r/r])^3\rangle=-12\epsilon r/d(d+2) \ .
\end{equation}

The general derivation of a Kolmogorov type relation for Eq.~(\ref{dyn}) proceeds along the same lines. We
consider the steady state condition $ \partial_t\langle q^a(0, t)q^a(\bm r, t)\rangle=0$ (here there is no summation over $a$). Employing the dynamical equation (\ref{dyn}) and using statistical symmetries, one finds
\begin{eqnarray}&&
0=\partial_t  \langle q^a(0, t)q^a(\bm r, t)\rangle=-2\frac{\partial}{\partial r_i}\langle q^a(0, t) F_i^a(\bm r, t)\rangle \nonumber\\&&
+2\langle q^a(0, t) f^a(\bm r, t)\rangle\nonumber
\\&&
-\frac{\partial }{\partial r_i}\left\langle q^a(0, t) \left(\sum_{jb}G^a_{i, jb}(\{\rho\}(\bm r, t))\nabla_j \rho^b(\bm r, t)\right)\right\rangle.
\nonumber
\end{eqnarray}
Again we consider the limit of large
correlation length $L$ of the forcing. Keeping all parameters of the system fixed we study the limit where $r$ in the above equation is large
but still much smaller than $L$. Because $r\ll L$ we have $f^a(\bm r, t)\approx f^a(0, t)$ and $\langle q^a(0, t) f^a(\bm r, t)\rangle\approx
\langle q^a(0, t) f^a(0, t)\rangle\equiv \epsilon$, where $\epsilon$ is some constant. Taking the limit of large $r$ we note that the last term
becomes less important as containing larger number of spatial derivatives. Hence we obtain
\begin{eqnarray}&&
\partial_i\langle q^a(0, t) F_i^a(\bm r, t)\rangle=\epsilon.
\end{eqnarray}
Assuming in addition isotropy one finds
\begin{eqnarray}&&
\langle q^a(0, t) F_i^a(\bm r, t)\rangle=\frac{\epsilon r_i}{d}, \label{general}
\end{eqnarray}
where $d$ is the space dimension.

In the following we will consider the implications of this derivation to compressible turbulence in relativistic hydrodynamics.
The implications to non-relativistic hydrodynamics are discussed in \cite{FFO}.
Consider first relativistic hydrodynamics in $(d+1)$-dimensional space-time,
with no conserved charges in addition to the stress-energy tensor. It is described by the energy density $\epsilon(x)$, the pressure $p(x)$ and the
$(d+1)$-velocity vector field $u^{\mu}(x), \mu=0,...,d$, satisfying
$u_{\mu}u^{\mu}=-1$. The stress-energy tensor obeys
\begin{equation}
\partial_{\nu}T^{\mu\nu}=0 \ ,
\label{cfteq}
\end{equation}
and the equations of relativistic hydrodynamics are determined by
the constitutive relation.
The constitutive relation
has the form of a series in the small parameter (Knudsen number)
\begin{equation}
Kn\equiv l_{cor}/L \ll 1 \ ,
\end{equation}
where $l_{cor}$ is the correlation length of the fluid (mean free path) and $L$ is
the scale of variations of the macroscopic fields.

 The constitutive relation reads
\begin{eqnarray}&&
T^{\mu\nu}(x)=\sum_{l=0}^{\infty}T^{\mu\nu}_l(x),\ \ T^{\mu\nu}_l\sim (Kn)^l, \label{series}
\end{eqnarray}
where $T^{\mu\nu}_l(x)$ is determined by the local values of
$u^{\mu},\epsilon, p$ and their derivatives of a finite order. Keeping
only the first term in the series gives ideal hydrodynamics, while
dissipative hydrodynamics arises when one keeps the first two terms
in the series.

The ideal hydrodynamics approximation for $T^{\mu\nu}$ does not
contain the spatial derivatives of the fields. The $l=0$ term in
(\ref{series}) gives the stress-energy tensor that reads
\begin{eqnarray}&&
T_{\mu\nu}= (\epsilon + p)u_{\mu}u_{\nu} + p \eta_{\mu\nu} \ ,
\label{ideal00}
\end{eqnarray}
where $\eta_{\mu\nu} = diag[-,+,+,..,+]$.

The dissipative hydrodynamics is obtained by keeping the $l=1$ term
in the series in Eq.~(\ref{series}), and includes
the shear and bulk viscosities. One has
an ambiguity in the form of the stress-energy tensor under a field
redefinition of the energy density, pressure and velocity that requires a choice of a frame.
However, the choice of the frame and the form of the viscous stress-tensor as well as the higher derivative terms
is of no importance in the following.

Consider the addition of random force term at large scale $L$ to  (\ref{cfteq}),
i.e.
\begin{equation}
\partial_{\nu}T^{\mu\nu}= f^{\mu} \ . \label{gen1}
\end{equation}

Following the same procedure as outlined above, with the assumptions made about the steady state regime at length scales $r$ much smaller
 than the force scale $L$ much larger than the viscous scale $l$,  one has the exact scaling relation
\begin{equation}
\langle T_{0j}(0, t) T_{ij}(\bm r, t)\rangle=\frac{\epsilon r_i}{d} \ ,
\label{stress}
\end{equation}
where $\langle T_{0j}(0, t) f_j(0, t)\rangle\equiv \epsilon$, and there is no summation over the index $j$.

As can be seen from the derivation, the exact result (\ref{stress}) is universal and does not depend on the frame or on the details on the
viscous and higher order terms.
It relies only on Eq.~(\ref{gen1}), which
holds for
general relativistic hydrodynamics (also independently of the possible existence of additional hydrodynamic equations such as
baryon number conservation). Therefor Eq.~(\ref{stress}) is a general scaling relation for any relativistic
hydrodynamics, where the hydrodynamic equations can be written as conservation laws.

Analogous relation holds for $\langle T_{00}(0, t) T_{0i}(\bm r, t)\rangle$ and corresponds to
the choice $q=T_{00}$ in Eq.~(\ref{general}).

A special case of the above is relativistic hydrodynamics of a conformal field theory (CFT).
In this case, the stress-energy tensor is traceless $T_{\mu}^{\mu} = 0$ and $\epsilon = d p \sim T^{d+1}$, where $T(x)$ is the
temperature field.
In the limit of
non-relativistic  macroscopic motions, the flow velocity $v$ being much smaller than the speed of light $c$,  the  relativistic conformal hydrodynamics equations
reduce to the non-relativistic incompressible Navier-Stokes
equations \cite{Fouxon:2008tb,Bhattacharyya:2008kq}.
Since the speed of sound in the relativistic conformal hydrodynamics is $v_s = \frac{c}{\sqrt{3}}$, the non-relativistic limit is
the limit of small Mach number.
In this limit, the exact scaling relation (\ref{stress}) reduces to the Kolmogorov relation (\ref{Kolmogorov}).
It is easy to see this explicitly,
by expanding $u^{\mu}=(\gamma, \gamma \bm v/c)$ and $\gamma=[1-v^2/c^2]^{-1/2}$ in $\frac{1}{c}$ and using
the expansion of the temperature as
\begin{eqnarray}&&
T=T_0\left[1+P/c^2+o(1/c^4)\right] \ ,
\label{expan}
\end{eqnarray}
where $P$ is the non-relativistic pressure.

We can generalize
further the above by considering charged hydrodynamics, where we have additional hydrodynamic global symmetry currents
conservation equation
\begin{equation}
\partial_{\mu} J^{\mu} = 0 \ .
\end{equation}
By adding a random force $f$ at large scale and the assumption of a steady state regime
we have the exact scaling relation
\begin{equation}
\langle J_{0}(0, t) J_{i}(\bm r, t)\rangle=\frac{\epsilon r_i}{d} \ ,
\label{JJ}
\end{equation}
where $\langle J_{0}(0, t) f(0, t)\rangle\equiv \epsilon$, and $f$ is an external field.

A particular interesting case is the hydrodynamics of anomalous relativistic gauge field theories, where the gauge invariant global symmetry current
is conserved at the classical level, but is not conserved  at the quantum level due to radiative corrections.
The anomaly coefficient $C$ depends on the details of the microscopic theory and it is interesting to know how this
coefficient is revealed in the hydrodynamic description  \cite{Son:2009tf} \footnote{Y.O. would like to thank S. Yankielowicz for discussions
on this issue.}.
In an external background electromagnetic field the anomalous equation reads
\begin{equation}
\partial_{\mu}T^{\mu\nu} = F^{\nu\lambda}J_{\lambda},~~~~~~~\partial_{\mu} J^{\mu} = C E^{\mu}B_{\mu} \ ,
\end{equation}
where $C$ is the anomaly coefficient and $E^{\mu} = F^{\mu\nu}u_{\nu}, B^{\mu} = \frac{1}{2}\epsilon^{\mu\nu\alpha\beta}u_{\nu}F_{\alpha\beta}$.
If we apply the same procedure as above where now the external electromagnetic field is a random external source, we see that the anomaly coefficient appears in the correlation function (\ref{JJ}).
Note, however, that in order to actually extract the anomaly coefficient we will probably need a knowledge of an
additional higher point correlation function in the steady state scaling regime.

Consider next the experimental implications of the above exact scaling relations.
The Reynolds number $Re$ for the relativistic hydrodynamics can be estimated  by the ratio of the ideal to viscous stress-energy tensors.
This gives
\begin{equation}
Re \sim \frac{T L}{\frac{\eta}{s}}  \ ,
\end{equation}
where $T$ is the temperature, $L$ is the characteristic length
scale, and $\eta/s$ is the ratio of the shear viscosity $\eta$ and the entropy density $s$.
A particularly interesting experimental setup is
that of relativistic heavy-ion collisions such as the RHIC program.
For gold collisions at RHIC, the characteristic scale $L$ is the radius of a
gold nucleus $L\sim 6$ Fermi, the temperature is the QCD scale
$T\sim 200$ MeV, and $\frac{\eta}{s} \sim \frac{1}{4\pi}$ is a characteristic value of strongly coupled gauge theories. With these one gets $R_e \gg 1$
 and one may expect an experimental realization of the steady state relativistic turbulence \cite{Romatschke:2007eb}.
It is therefore of much interest to see whether there exists in the RHIC data an experimental signature
of the scaling relation (\ref{stress}).

The connection between the hydrodynamic description and the experimentally observed particles after hadronization
is made by the requirement that the hydrodynamic and kinetic theory stress-energy tensors are the same at freeze-out.
The latter stress-energy tensor is expressed in terms of the one-particle distribution function and is related
to the measured angular distribution of the particles.
Thus, one may expect that the hydrodynamic correlation functions of the stress-energy tensor
will be reflected in the measured particle data.

Another experimental setup, where one may be able to study universal properties of relativistic turbulence
is condensed matter physics.
For instance, it has been noticed in \cite{CM} that there is an
emergent relativistic symmetry of electrons in graphene near its quantum critical point, for which a relativistic
nearly ideal fluid description may be appropriate \cite{MM}.

Much like the numerical evaluation of anomalous exponents in the scaling regime of non-relativistic incompressible fluids, one can
attempt to study the existence of a scaling regime of relativistic hydrodynamics via numerical simulations. It would be interesting to use the existing relativistic hydrodynamic simulations for heavy ion collisions, where one can vary the system parameters such as the nucleus size $L$, in order to study
the scaling relations at high Reynolds number.
This requires the introduction of a random force or random initial conditions and a tuning to the scaling
regime, where the viscous and higher order terms are negligible.

A large class of compressible and incompressible hydrodynamic systems have a geometrical realization as the dynamics of black hole
event horizons
in asymptotically anti-de-Sitter space \cite{Bhattacharyya:2008xc,Eling:2009pb,Eling:2009sj}.
In these descriptions, the forces arise from adding additional bulk fields, such as a dilaton,
to Einstein gravity.
It would be interesting to construct the geometrical analog of the above derivation of exact scaling relations.

\section*{Acknowledgements}

We would like to thank U. Wiedemann
for a discussion.
The work is supported in part by the Israeli
Science Foundation center of excellence, by the Deutsch-Israelische
Projektkooperation (DIP), by the US-Israel Binational Science
Foundation (BSF), and by the German-Israeli Foundation (GIF).


\begin{thebibliography} {99}

\bibitem{Kolmogorov} A. N. Kolmogorov, Dokl. Akad. Nauk SSSR, \textbf{32} 16 (1941).

\bibitem{Frisch} U. Frisch, \textit{Turbulence: The Legacy of A. N.
Kolmogorov}, Cambridge University Press, 1995.

\bibitem{Gawedzki} K. Gawedzki, {\it Easy turbulence}, chao-dyn/9907024.

\bibitem{FalkovichSreenivasan} G. Falkovich and K. R. Sreenivasan, Phys. Today, \textbf{59} (4), 43 (2006).

\bibitem{Yaglom} A. M. Yaglom, Dokl. Akad. Nauk. SSSR, \textbf{69}, 743 1949.

\bibitem{Chandrasekhar} S. Chandrasekhar, Proc. Roy. Soc. A \textbf{204}, 435 (1951).

\bibitem{PouquetPolitano} H. Politano and A. Pouquet, Phys. Rev. E \textbf{57}, R21 (1998).

\bibitem{Galtier} S. Galtier, Phys. Rev. E \textbf{77}, 015302(R) (2008).

\bibitem{FFO} G. Falkovich, I. Fouxon and Y. Oz, J. Fluid. Mech. \textbf{644}, 465 (2010),arXiv:0909.3404 [chao-dyn].



\bibitem{Forster} D. Forster, \textit{Hydrodynamic Fluctuations, Broken Symmetry, And Correlation Functions},
Perseus Books, 1975.

%\bibitem{Kritsuk} A. G. Kritsuk, M. L. Norman, P. Padoan, and R. Wagner, Astrophys. J. \textbf{665} (1), 416 (2007).



%\cite{Fouxon:2008tb}
\bibitem{Fouxon:2008tb}
  I.~Fouxon and Y.~Oz,
  %``Conformal Field Theory as Microscopic Dynamics of Incompressible Euler and
  %Navier-Stokes Equations,''
  Phys.\ Rev.\ Lett.\  {\bf 101}, 261602 (2008),
 [arXiv:0809.4512 [hep-th]].

%\cite{Bhattacharyya:2008kq}
\bibitem{Bhattacharyya:2008kq}
  S.~Bhattacharyya, S.~Minwalla and S.~R.~Wadia,
  %``The Incompressible Non-Relativistic Navier-Stokes Equation from Gravity,''
  arXiv:0810.1545 [hep-th].
  %%CITATION = ARXIV:0810.1545;%%



%\cite{Son:2009tf}
\bibitem{Son:2009tf}
  D.~T.~Son and P.~Surowka,
  %``Hydrodynamics with Triangle Anomalies,''
  Phys.\ Rev.\ Lett.\ {\bf 103}, 191601 (2009),
  arXiv:0906.5044 [hep-th].
  %%CITATION = ARXIV:0906.5044;%%


%\cite{Romatschke:2007eb}
\bibitem{Romatschke:2007eb}
  P.~Romatschke,
  %``Fluid turbulence and eddy viscosity in relativistic heavy-ion collisions,''
  Prog.\ Theor.\ Phys.\ Suppl.\  {\bf 174}, 137 (2008).
%  [arXiv:0710.0016 [nucl-th]].
  %%CITATION = PTPSA,174,137;%%


%\bibitem{WIP} Work in progress.

\bibitem{CM} D. Sheehy and J. Schmalian,
Phys.\ Rev.\ Lett.\ {\bf 99}, 226803 (2007).

\bibitem{MM} M. Muller, in "AdS/CFT: strongly coupled systems and exact results", IHP/ENS Paris 26/27 (2009).


\bibitem{Bhattacharyya:2008xc}
  S.~Bhattacharyya {\it et al.},
  %``Local Fluid Dynamical Entropy from Gravity,''
  JHEP {\bf 0806}, 055 (2008),
  [arXiv:0803.2526 [hep-th]].

%\cite{Eling:2009pb}
\bibitem{Eling:2009pb}
  C.~Eling, I.~Fouxon and Y.~Oz,
  %``The Incompressible Navier-Stokes Equations From Membrane Dynamics,''
  Phys.Lett.B {\bf 680}, 496 (2009),
  arXiv:0905.3638 [hep-th].
  %%CITATION = ARXIV:0905.3638;%%


%\cite{Eling:2009sj}
\bibitem{Eling:2009sj}
  C.~Eling and Y.~Oz,
  %``Relativistic CFT Hydrodynamics from the Membrane Paradigm,''
  JHEP {\bf 1002}, 069 (2010),
arXiv:0906.4999 [hep-th].
  %%CITATION = ARXIV:0906.4999;%%



\end{thebibliography}
\end{document}